\documentclass[a4paper,11pt]{article}
\usepackage{pos}
\usepackage{lineno}
\usepackage{natbib}
\usepackage{multirow}
\newcommand{\gm}{\ensuremath{\gamma}}

\newcommand{\swift}{\textit{Swift}}
\newcommand{\integral}{\textit{INTEGRAL}}
\newcommand{\xmm}{\textit{XMM-Newton}}

\newcommand\Bstrut{\rule[-0.9ex]{0pt}{0pt}}  % = 'bottom' strut

\begin{document}
%\linenumbers
\title{Multi-wavelength study of Mrk 421 during a TeV flare}
%% \ShortTitle{...}
\author*[a,b]{A.~Gokus}
\author[b]{A.~Kreikenbohm}
\author[b]{K.~Leiter}
\author[c,d,\#]{T.~Bretz}
\author[a]{T.~Dauser}
\author[b,\#]{D.~Dorner}
\author[e,\#]{D.~Elsaesser}
\author[b]{F.~Eppel}
\author[b]{J.~He{\ss}d{\"o}rfer}
\author[b]{M.~Kadler}
\author[f]{A.~Kraus}
\author[g]{M.~Kreter}
\author[a]{I.~Kreykenbohm}
\author[b]{M.~Langejahn}
\author[b,\#]{K.~Mannheim}
\author[a]{P.~Thalhammer}
\author[a]{J.~Wilms}
\author[c]{and the FACT collaboration: A.~Arbet-Engels}
\author[e]{D.~Baack}
\author[h]{M.~Balbo}
\author[c]{A.~Biland}
\author[e]{J.~Buss}
\author[b]{L.~Eisenberger}
\author[c]{D.~Hildebrand}
\author[b]{R.~Iotov}
\author[b]{A.~Kalenski}
\author[c]{A.~Mitchell}
\author[c]{D.~Neise}
\author[e]{M.~Noethe}
\author[b]{A.~Paravac}
\author[e]{W.~Rhode}
\author[b]{B.~Schleicher}
\author[h]{V.~Sliusar}
\author[h]{R.~Walter}
\affiliation[a]{Dr. Karl Remeis-Observatory and Erlangen Centre for Astroparticle Physics, Universit\"at Erlangen-N\"urnberg, Sternwartstr.~7, 96049 Bamberg, Germany}
\affiliation[b]{Lehrstuhl f\"ur Astronomie, Universit\"at W\"urzburg, Emil-Fischer-Stra{\ss}e 31, 97074 W\"urzburg, Germany}
\affiliation[c]{ETH Z\"urich, Institute for Particle Physics and Astrophysics, Otto-Stern-Weg 5, 8093 Z\"urich, Switzerland}
\affiliation[d]{RWTH Aachen, Experimental Physics 3a, Sommerfeldstra{\ss}e, 52074 Aachen, Germany}
\affiliation[e]{TU Dortmund, Experimental Physics 5, Otto-Hahn-Str. 4a, 44227 Dortmund, Germany}
\affiliation[f]{Max-Planck-Institut für Radioastronomie, Auf dem Hügel 69, 53121, Bonn, Germany}
\affiliation[g]{Centre for Space Research, North-West University, Potchefstroom, 2520, South Africa}
%\affiliation[f]{Department of Physics, Pennsylvania State University, University Park, PA 16802, USA}
\affiliation[h]{University of Geneva, Department of Astronomy, Chemin d\'Ecogia 16, 1290 Versoix, Switzerland}
\affiliation[\#]{also in FACT}
\emailAdd{andrea.gokus@fau.de}

\abstract{
The blazar Mrk\,421 shows frequent, short flares in the TeV energy regime.
Due to the fast nature of such episodes, we often fail to obtain sufficient simultaneous information about flux variations in several energy bands. To overcome this lack of multi-wavelength (MWL) coverage, especially for the pre- and post-flare periods, we have set up a monitoring program with the FACT telescope (TeV energies) and the Neil Gehrels \swift~Observatory (X-rays).
On 2019 June 9, Mrk 421 showed a TeV outburst reaching a flux level of more than two times the flux of the Crab Nebula at TeV energies.
We acquired simultaneous data in the X-rays with additional observations by \xmm~and \integral. For the first time, we can study a TeV blazar in outburst taking advantage of highly sensitive X-ray data from \xmm~and \integral~combined. Our dataset is complemented by pointed radio observations by Effelsberg at GHz frequencies.
We present our first results, including the \gm-ray and X-ray light curves, a timing analysis of the X-ray data obtained with \xmm~, as well as the radio spectra before, during and after the flare.
}

\FullConference{37$^{\rm{th}}$ International Cosmic Ray Conference (ICRC 2021)\\
		July 12th -- 23rd, 2021\\
		Online -- Berlin, Germany}
%% \tableofcontents
\maketitle
\section{Introduction}\vspace{-0.2cm}
The active galactic nucleus (AGN) Mrk\,421 is one of the closest blazars with a redshift of $z=0.031$ \cite{ulrich1975}. It belongs to the group of high-peaked BL Lac type objects, meaning the low-energy hump in its typical two-hump-shaped spectral energy distribution (SED) peaks above $10^{15}$\,Hz and can show extreme-blazar properties during flares \cite[e.g.][]{Sahu2021}. Its optical spectrum is strongly dominated by continuum emission, unlike the optical spectrum of flat-spectrum radio quasars, in which emission lines with a width of $>5$\AA~can be found.
While the low-energy hump of a blazar's SED is caused by synchrotron emission by leptons, the origin of the high-energy hump is still under debate. One scenario is that the highest energies are also produced by leptons via Inverse Compton (IC) processes, while the other scenario includes protons, which can either produce synchrotron radiation or interact with photons, which creates particles that scatter down to high-energetic leptons. % hier fehlen noch referenzen, evtl. auch weglassen
Due to its proximity, Mrk\,421 is one of the best studied sources, but we are still far from fully understanding the underlying physics of the observed phenomena across the electromagnetic spectrum. Detailed  multiwavelength (MWL) observations during well-defined, characteristic emission states hold the potential to yield key observational data but are logistically demanding and challenging to set up.

Mrk\,421 has been a known TeV \gm-ray emitter since its detection by the Whipple telescope \cite{punch1992}, and frequently shows enormous flux variations (see, e.g., \cite{dorner2019}), exceeding a flux of three times the flux of the Crab Nebula at TeV energies (Crab Units, CU; see, e.g.,\cite{macomb1995},\cite{albert2007},\cite{bartoli2016}), as well as a correlation between optical and GeV emission \cite{carnereo2017} during flares.
In addition, Aleksic et al. (2015) found a correlation between TeV and X-ray emission during non-flaring emission states with zero time lag \cite[][]{aleksic2015}. Because the correlation persists both in flaring and non-flaring time ranges, these authors conclude that it is difficult to explain the observed long-term variability of Mrk\,421 within hadronic scenarios and therefore favour leptonic scenarios.

Mrk\,421 belongs to a sample of TeV emitters that show particularly slow apparent jet speeds as measured by Very Long Baseline Interferometry (VLBI) observations in the centimeter radio band. Its viewing angle $\alpha$ has been constrained to $2^{\circ}<\alpha<5^{\circ}$ using VLBA data \cite{blasi2013}, and the fastest measured jet speeds range from $0.09$\,c \cite{piner2010} to $0.4$\,c \cite{kellermann2004}. The physical conundrum here is that the most energetic blazars seem to show relatively low jet bulk motion indicative of only mildly relativistic plasma, which contradicts the extreme \gm-ray properties. This problem had been coined the \textit{Doppler crisis}. As a solution, Hervet et al. (2019) propose that the observed knots are actually recollimation shocks in which particles get accelerated along the jet \cite{hervet2019}. As a test target, they chose Mrk\,421, and studied its 13 years long X-ray light curve, obtained with \swift~. They probed for an intrinsic variability pattern caused by particles moving through each radio knot, which they found, but a similar study of the TeV light curve for Mrk\,421 is necessary to establish their theory.

In this work, we present TeV, X-ray, and radio data from Mrk\,421, which were taken during a flare of the source in June 2019. The flare had been detected by the Cherenkov telescope FACT, and MWL follow-up observations had been performed with the Neil Gehrels \swift~Observatory (\swift), \xmm, \integral~, and the 100m-dish radio telescope at Effelsberg. The success of the rapid follow-up observations is owed to our MWL program, which has been planned in great detail and run since 2012 \cite{Kreikenbohm2019}. The setup of this MWL program  will be presented in this work as well.

\section{Monitoring program}\vspace{-0.2cm}
In order to fully comprehend the emission of blazars, and given their very variable nature, it would be desirable to obtain long-term uninterrupted data sets, that cover the full electromagnetic spectrum. As this is logistically not feasible, we have set up a program to combine a long-term snapshot monitoring of selected blazars with deep Target-of-Opportunity (ToO) short-term quasi-simultaneous MWL observations during characteristic flaring states.
In this program, we make use of the Cherenkov telescope FACT and the X-ray satellite \swift~to monitor the flux of Mrk\,421 at TeV (with nightly cadence), and X-ray and optical/UV energies (with 4-5 day cadence), respectively. In addition, we defined ToO criteria to trigger further instruments for full MWL coverage, including \integral~, \xmm, and the 100-m radio telescope in Effelsberg. The program has been run since 2012 and first results from a first moderately bright flare of Mrk\,421 have been presented by Kreikenbohm (2019) \cite{Kreikenbohm2019}.

\begin{figure}[t]
   \centering
%   \begin{minipage}[c]{0.65\textwidth}
    \includegraphics[width=\textwidth]{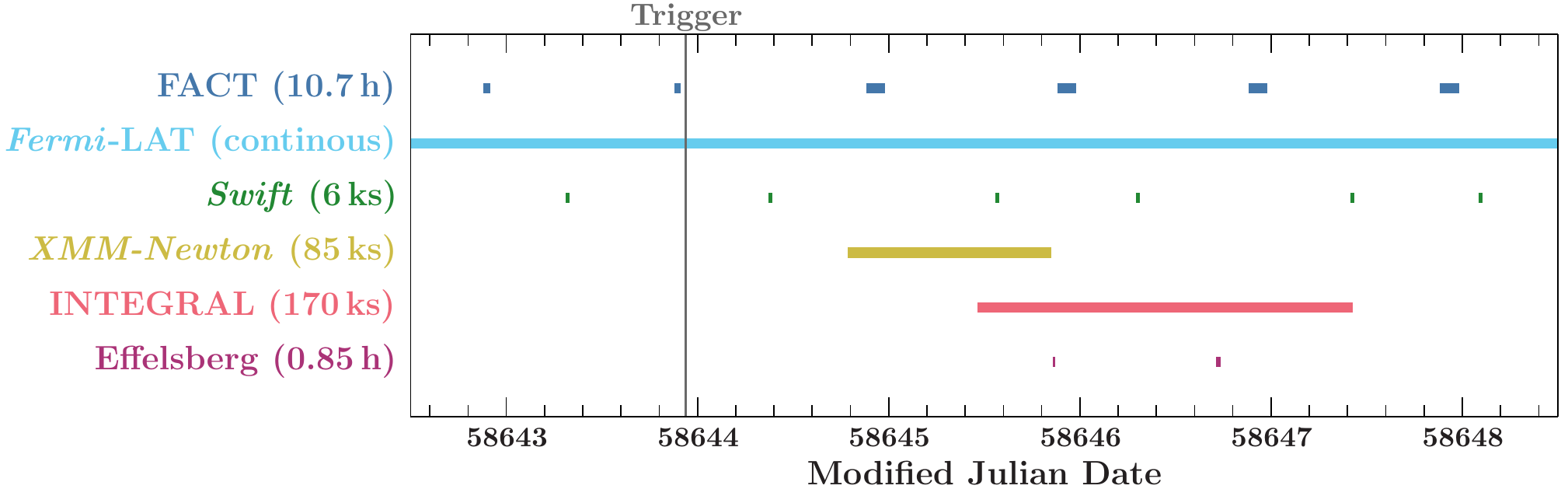}
%    \end{minipage}\hfill
%    \begin{minipage}[c]{0.33\textwidth}
      \caption{Overview of the follow-up observation of our triggered multi-wavelength campaign between June 8 and 14, 2019. Exposure times refer to the accumulated values during the time range displayed.}
         \label{fig:monitoring_overview}
%    \end{minipage}
\end{figure}

On 2019 June 9, FACT registered a TeV flux brighter than 2~CU, and we sent out a trigger to \swift, \xmm, and \integral~ at 2019 June 9 22:30 UTC. Daily \swift~pointings were requested to obtain a dense monitoring at X-ray energies during and after the flare, while a deep pointing with \xmm~was performed to allow a detailed variability study on time scales smaller than one day. Because \swift~and \xmm~cover the X-ray energy range from roughly $0.5-10$\,keV, the \integral~mission is included in our program to gain data for the harder X-ray range from 15\,keV up to 10\,MeV, as well.
Figure~\ref{fig:monitoring_overview} shows the timeline of our MWL observations from 2019 June 8, 12:00 UTC until 2019 June 14, 12:00 UTC, with a true representation of the observation times in relation to each other.

\section{Instruments and Data}\vspace{-0.3cm}
\subsection{The First G-APD Cherenkov Telescope}\vspace{-0.2cm}
The First G-APD Cherenkov Telescope (FACT) observes the \gm-ray sky at energies above 300\,GeV since October 2011 \cite{anderhub2013}. Its major objective is to monitor a small sample of known TeV-emitting blazars continuously in order to obtain an unbiased, long-term view on the gamma-ray activity of these sources. During this campaign in 2019, FACT has observed Mrk\,421 for about 570~hours of which 10.7~hours are presented here. Based on the monitoring of FACT, the trigger to the X-ray satellites \xmm~and \integral~was sent. The data were analyzed using the Modular Analysis and Reconstruction Software (MARS) \cite{2010apsp.conf..681B}. In the context of the database-based analysis \cite{database-based}, the 'light curve cuts' as described in \cite{2019ICRC...36..630B} were used for background suppression. The data quality selection was done based on the artificial trigger rate \cite{2019ICRC...36..630B,2017ICRC...35..779H}. More details on the analysis are available in \cite{2021A&A...647A..88A}.  

\subsection{The Neil Gehrels \swift~Observatory}\vspace{-0.2cm}
The Neil Gehrels \swift~Observatory has three instruments on-board: the UV/Optical Telescope (UVOT), the X-ray Telescope (XRT), and the Burst Alert Telescope (BAT) \cite{gehrels2004}. In this work, we only present the data from \swift/XRT, which observes X-rays in an energy range between 0.5\,keV and 10\,keV, and focus on the analysis of observations between MJD 58636 and MJD 58652. Each of these observations was taken in the window timing mode, and with a net exposure time of 1\,ks. We extracted the spectra using the \texttt{xrtpipeline} available within HEASOFT (v6.26).
We fit all spectra with an absorbed powerlaw, for which the influence of photoelectric absorption was frozen to a value derived from the measured column of neutral hydrogen N$_{\mathrm{H}}=0.0133$ \cite{hi4pi_2016}. The X-ray fluxes of Mrk\,421 around the time of the flare are plotted in Fig.~\ref{fig:lc_xray_gray}.
%value of the hydrogren column was frozen to the Galactic value N$_{\mathrm{H}}=0.0133$ \cite{hi4pi_2016}.
%\texttt{MK: More details and/or a reference to the detailed procedure.}

\subsection{The X-ray satellite \xmm}\vspace{-0.2cm}
The \xmm~satellite carries three Wolter X-ray telescopes as well as an optical telescope. X-ray spectra are taken with three cameras and different CCD chips \cite{jansen2001}. The observation on Mrk\,421 on 2019 June 10 (ObsID: 0845000901) was taken in timing mode, however, because of the immense flux increase during the observation, it was necessary to change the observational mode to burst mode. The data of the EPIC-pn camera was reduced using standard methods\footnote{https://www.cosmos.esa.int/web/xmm-newton/sas-thread-pn-spectrum-timing} of the \xmm~Science Analysis System (Version 18.0.0).
%\texttt{MK: Add an example reference where this is described in more detail.}
Because of pile-up in timing mode, we excluded the inner five columns (RAWX = 36-40) of the chip. 
We extract a light curve with a time binning of 100\,s, which is shown in Fig.~\ref{fig:xmm_psd} (left). The count rates measured for Mrk\,421 are the highest rates ever seen for this source by \xmm. In addition, we extract time-resolved spectra with an exposure of 3\,ks each, and fit these spectra with the same model and N$_{\mathrm{H}}$ value as the \swift~spectra. The resulting flux values are shown in the \xmm~light curve in Fig.~\ref{fig:lc_xray_gray}.

\subsection{The Effelsberg 100-m radio telescope}
We used the 100m-dish radio telescope in Effelsberg to observe Mrk\,421 at a broad range of frequencies\footnote{Secondary receiver: 4.85, 10.45, 14.25, 16.75, 19.25, 21.15, 22.85, 24.75, 35.75, and 38.25\,GHz} across the cm band before (May 5), during (June 7, 11, 12), and after (July 5, November 4, 16) the TeV flare. 
The observations have been performed as cross scans in azimuth and
elevation over the (unresolved) source. Pointing offsets were corrected and individual subscans were averaged. After corrections for the atmosphere’s opacity and the gain-elevation effect, the flux densities were calibrated using suitable calibrators like 3C286 \cite[see, e.g.,][]{Kraus2003}.

\section{Variability analysis}\vspace{-0.2cm}
\begin{figure}[t]
   \centering
    \includegraphics[width=\textwidth]{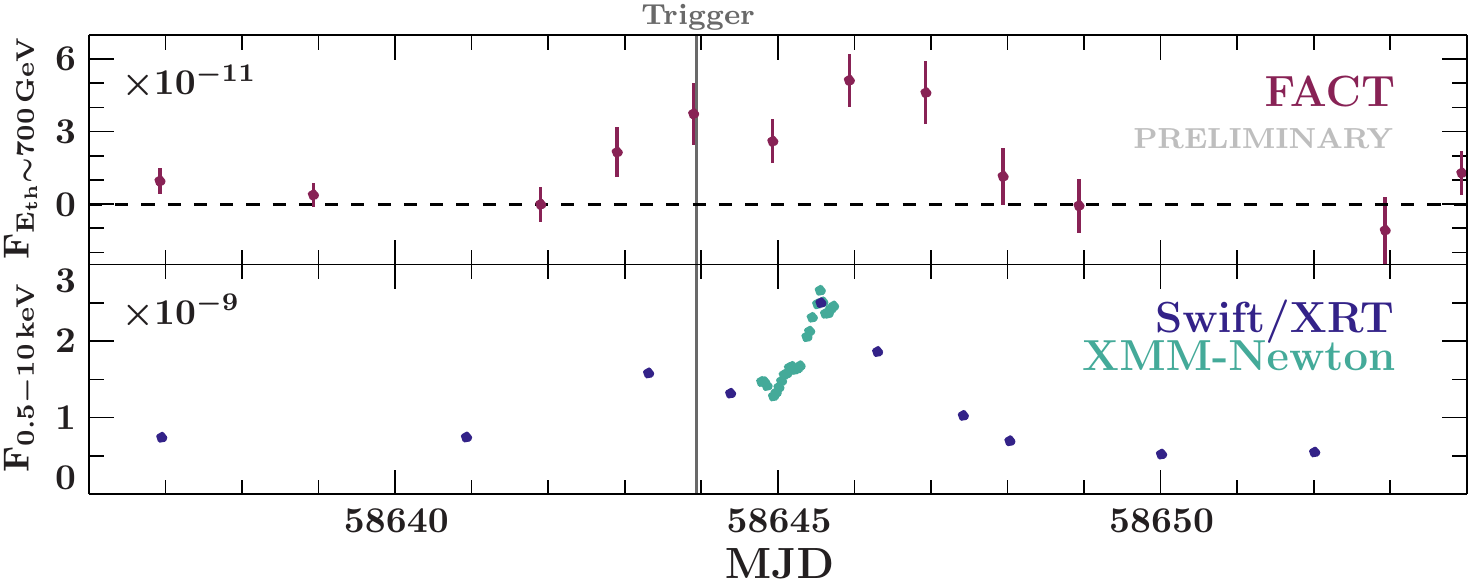}
     \caption{X-ray and \gm-ray light curves of Mrk\,421 between 2019 June 2 (MJD~58636) and 2019 June 20 (MJD~58654). The \gm-ray flux unit is given in ph\,cm$^{-2}$\,s$^{-1}$, and the X-ray flux units are given in erg\,cm$^{-2}$\,s$^{-1}$. The trigger time for follow-up observation is marked by the grey vertical line.}
         \label{fig:lc_xray_gray}
\end{figure}
Here, we present the \gm-ray light curve for energies above 700\,GeV, obtained with FACT, and the X-ray light curve, composed of \swift~and \xmm~pointings. A thorough variability analysis, also including other wavelengths, will be presented in a later work.

The \gm-ray and X-ray light curves are shown in Fig.~\ref{fig:lc_xray_gray}. We additionally mark the time, when we sent a trigger to other telescopes for MWL follow-up observations. 
After the trigger, the source remained in a high TeV state.
Two and three nights later, respectively, the \gm-ray flux even exceeded the flux at the time of the trigger. Due to the dense \swift~monitoring and the long \xmm~observation, we were able to track the behaviour in the X-ray energy regime in unprecedented detail. In the X-rays, the highest flux for the depicted time range is measured at $\sim$MJD~58645.5. After that, the flux seems to decrease over about three days until the flux is at the same level as before the flare.

\begin{figure}[t]
   %\centering
    \begin{minipage}[c]{0.62\textwidth}
      \includegraphics[width=.99\textwidth]{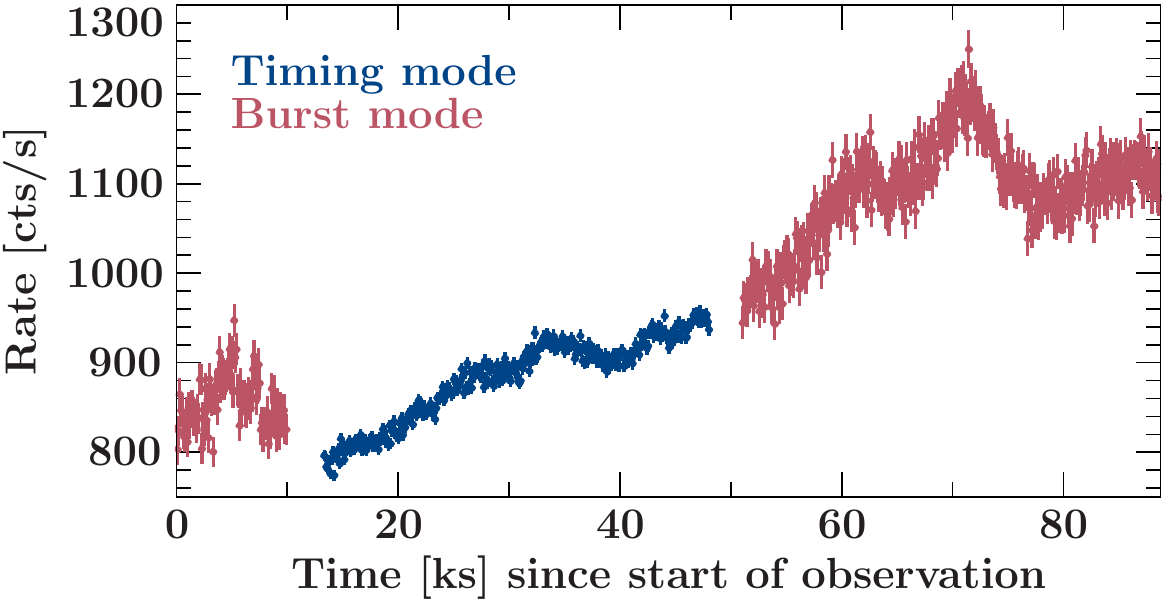}
    \end{minipage}\hfill
   \begin{minipage}[c]{0.36\textwidth}
    \includegraphics[width=.99\textwidth]{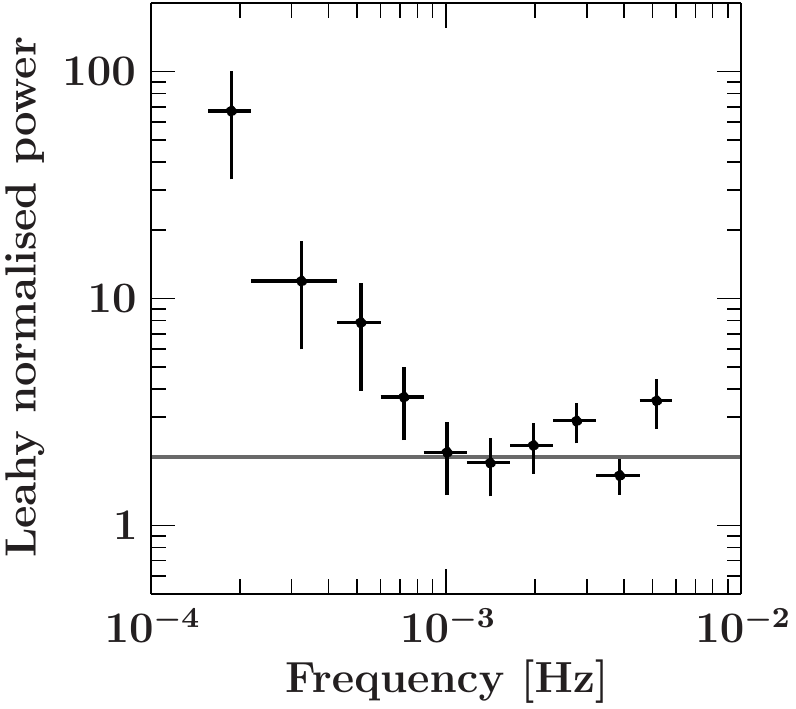}
    \end{minipage}
    \caption{\textit{Left:} \xmm~ light curve with a bin size of 100s. Data taken in timing and burst mode are shown in blue and red, respectively. \textit{Right:} Power spectral density computed from the \xmm~light curve with 100s binning. The horizontal line marks the level of Poisson noise.}
     \label{fig:xmm_psd}
\end{figure}

We compute the power spectral density (PSD) from the 100s-binned \xmm~ light curve, yielding a Nyquist frequency $f_{\mathrm{Ny}}=5\cdot10^{-3}$ Hz. We use the Leahy normalisation in order to directly distinguish Poisson noise from source intrinsic variability. The resulting PSD is shown in Fig.~\ref{fig:xmm_psd}. At a frequency of $10^{-3}$\,Hz ($\sim16$\,minutes), the power reaches the level of the Poisson noise.

\begin{figure}[t]
%   \centering
    \begin{minipage}[c]{0.73\textwidth}
    \includegraphics[width=.99\textwidth]{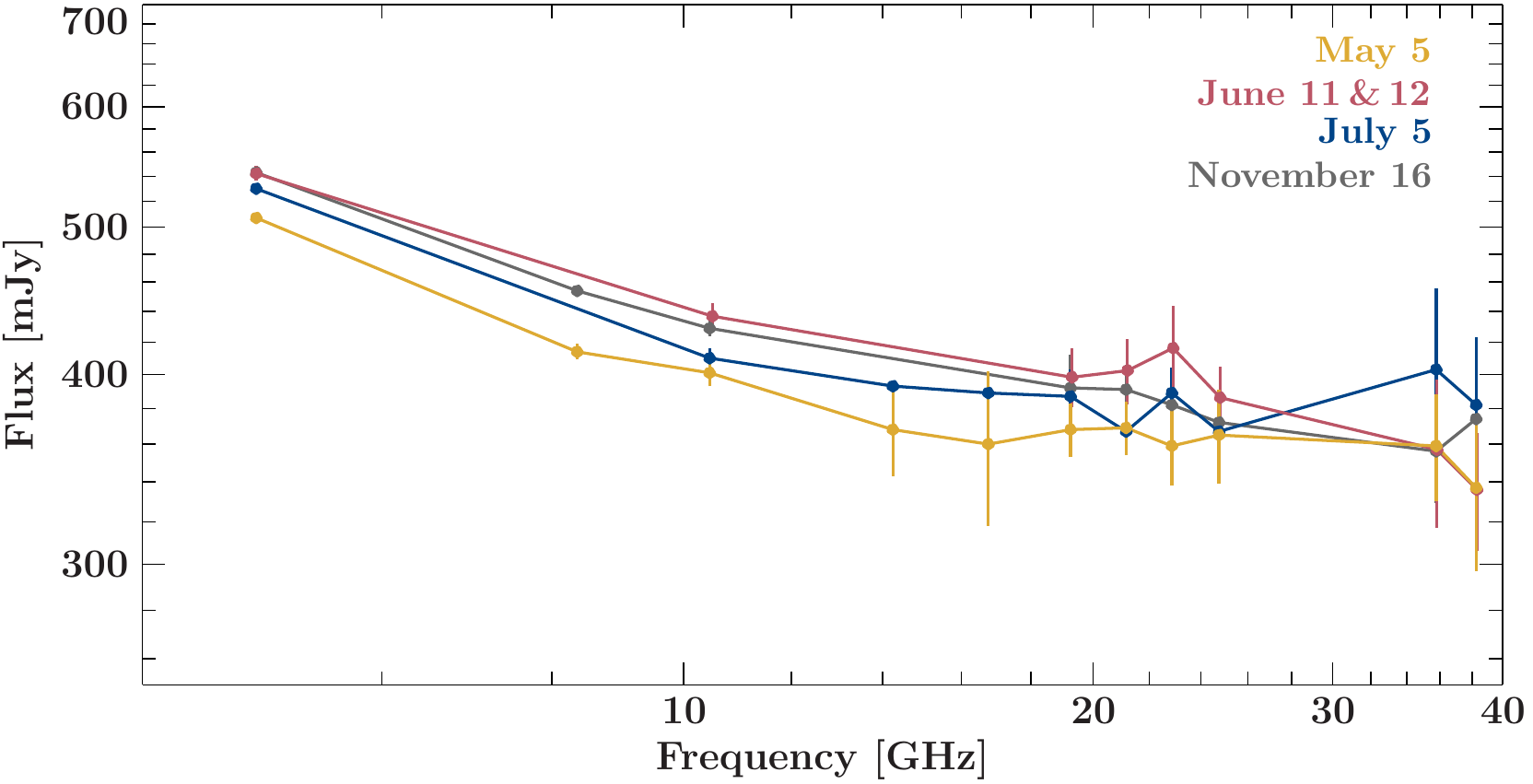}

    \end{minipage}\hfill
    \begin{minipage}[c]{0.26\textwidth}
         \caption{Radio spectra of Mrk\,421 before, during and after the TeV flare in June. The observations on June 11 and 12 have been conducted at different frequencies and were therefore combined into one spectrum.}
          \label{fig:radio_spectra}
    \end{minipage}

\end{figure}

\section{Radio spectra with the Effelsberg 100-m radio telescope}
In Fig.~\ref{fig:radio_spectra}, we show radio spectra taken with the 100m-dish radio telescope Effelsberg between 4.85\,GHz and 38.25\,GHz before, during, and after the TeV flare. In between these times, no prominent spectral changes can be observed, except for a slight flattening on July 5, which is $\approx1$ month after the TeV and X-ray flare.
%which is caused by a higher flux at 8mm (35.75 and 38.25\,GHz) compared to larger wavelengths on that day. 
The spectrum for June 11/12 exhibits the highest overall flux, but these variations are well within the typical range that can be observed for Mrk\,421 (see, e.g., \cite{kadler2021} in these proceedings for a radio light curve of Mrk\,421 in the 14mm and 7mm bands, taken as part of the TELAMON program). This shows that the \gm-ray flare has not caused a major response of the parsec-scale radio jet as probed on a cadence of  5 months after the event.

\section{Summary and Outlook}\vspace{-0.2cm}
%Although being the closest blazar, Mrk\,421 is still a very interesting source to continuously study as many things are not yet understood. 
We have set up a sophisticated MWL program to fully characterize a blazar flare including pre-flare information, quasi-simultaneous MWL data during the flare and long-term follow-up observations.
The combination of dense X-ray and VHE \gm-ray monitoring with pre-approved trigger opportunities at facilities across the electromagnetic spectrum, has yielded a unique MWL data set of a flare in the closest blazar Mrk\,421 in June 2019, detected by FACT. The high activity at \gm-rays lasted $\approx4$ days, with the highest flux being measured on June 11, which seems to coincide with the highest X-ray flux.
During the \xmm~observation, we find sub-hour variability in the X-rays. The pc-scale radio jet has not shown any major response to the high-energy activity.
In a later work, we will perform a more thorough analysis of the MWL data, including a deeper analysis of the behaviour at X-ray energies 
and in the TeV band.
%with regard to spectral hardening and a possible shift of the synchrotron peak.

\acknowledgments\vspace{-0.4cm}
{\small A. Gokus was funded by the Bundesministerium für Wirtschaft und Technologie under Deutsches Zentrum für Luft- und Raumfahrt (DLR grant number 50OR1607O) and by the German Science Foundation (DFG grant number KR 3338/4-1).
The important contributions from ETH Zurich grants ETH-10.08-2 and ETH-27.12-1 as well as the funding by the Swiss SNF and the German BMBF (Verbundforschung Astro- und Astroteilchenphysik) and HAP (Helmoltz Alliance for Astroparticle Physics) to the FACT project are gratefully acknowledged. Part of this work is supported by Deutsche Forschungsgemeinschaft (DFG) within the Collaborative Research Center SFB 876 "Providing Information by Resource-Constrained Analysis", project C3. We are thankful for the very valuable contributions from E. Lorenz, D. Renker and G. Viertel during the early phase of the FACT project. We thank the Instituto de Astrofísica de Canarias for allowing us to operate the telescope at the Observatorio del Roque de los Muchachos in La Palma, the Max-Planck-Institut für Physik for providing us with the mount of the former HEGRA CT3 telescope, and the MAGIC collaboration for their support.
This work was partly based on observations with the 100-m telescope of the MPIfR (Max-Planck-Institut für Radioastronomie) at Effelsberg.
We made use of a collection of ISIS functions (ISISscripts) provided by ECAP/Remeis observatory and MIT (http://www.sternwarte.uni-erlangen.de/isis/).
The colours used in the Figures are taken from Paul Tol’s colour schemes and templates (https://personal.sron.nl/ pault/).}
% Bibliography
%\newpage
\bibliographystyle{JHEP}
\bibliography{references}
\end{document}